\RequirePackage{fix-cm}
\documentclass[smallextended]{svjour3}       
\smartqed  
\usepackage{graphicx}
\usepackage{float}
\usepackage{algpseudocode}
\usepackage{qtree}
\usepackage{newfloat}

\DeclareFloatingEnvironment[fileext=lod]{diagram}
\begin{document}

\title{Using modular decomposition technique to solve the maximum clique problem}

\author{Irina Utkina 
}

\institute{I. Utkina \at
              Laboratory of Algorithms and Technologies for Network Analysis, Higher School of Economics in Nizhni Novgorod, 136, Rodionova Str., N.Novgorod, Russia \\
              \email{iutkina@hse.ru}
}

\date{Received: date / Accepted: date}

\maketitle

\begin{abstract}
In this article we use the  modular decomposition technique for exact solving the weighted maximum clique problem. Our algorithm takes the modular decomposition tree from the paper of Tedder et. al. and finds solution recursively. Also, we propose algorithms to construct graphs with modules. We show some interesting results, comparing our solution with Ostergard’s algorithm on DIMACS benchmarks and on generated graphs
\end{abstract}

\section{Introduction}
Today graphs can be used in variety fields, such as biology, chemistry\cite{bib1,bib4}, data analysis, mathematics and others, as a structure of data. Due to enormous expand of information, the size of graphs for analysis is increasing, it can be hundred, thousand and even hundred of thousands vertices. Since the computational time of any algorithm on graphs depends on its size, it has become a great problem for community. There are many graph decomposition techniques to reduce a graph to its smaller fragments; one of them is modular decomposition. In this article we use the fastest algorithm for constructing modular decomposition proposed by Tedder et. al\cite{bib2}. It creates a modular decomposition tree for any input graphs in linear time. Then we use this tree to solve the maximum clique problem on graphs from DIMACS benchmarks using Ostergard’s\cite{bib5} algorithm, and compare computational time with Ostergard’s algorithm without modular decomposition technique. Also, we construct some other types of graphs, such as co-graphs and graphs of mutual simplicity.
\section{Modular decomposition algorithm}
\begin{figure}
\includegraphics[width=\textwidth]{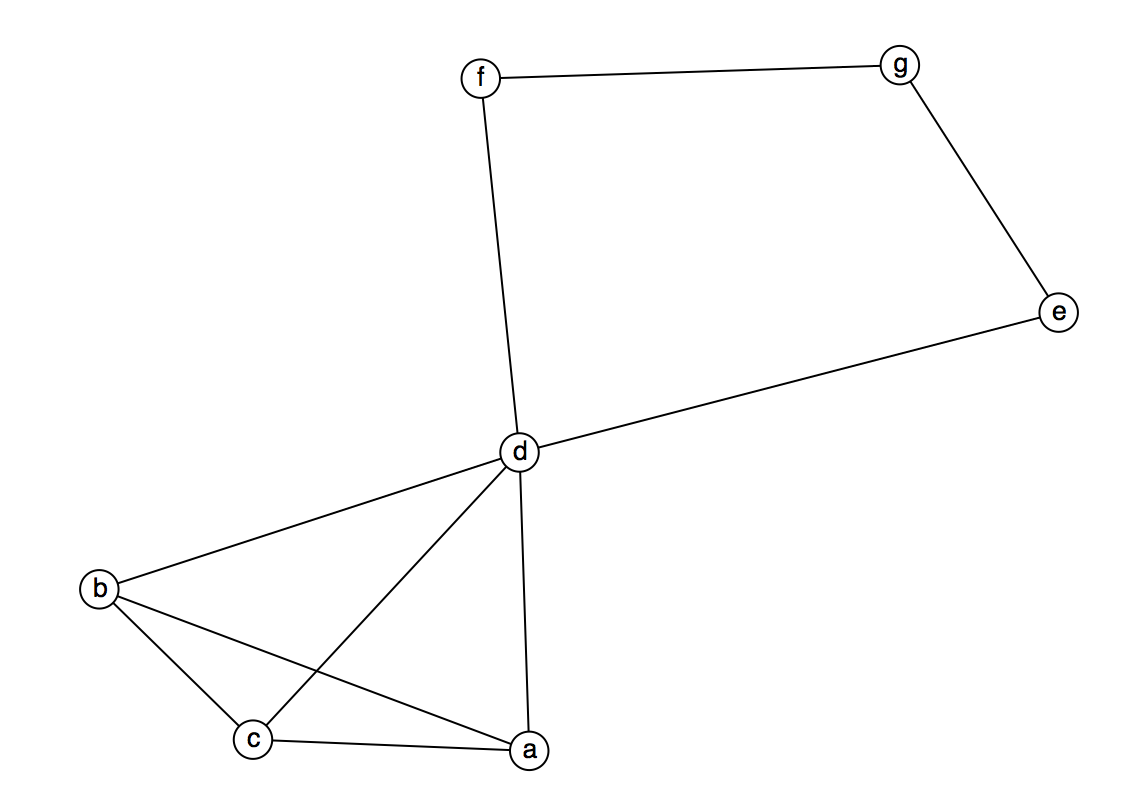}
\caption{Example 1\cite{bib6}\newline}
\label{lab:1}
\end{figure}

\begin{figure}
\includegraphics[scale=0.5]{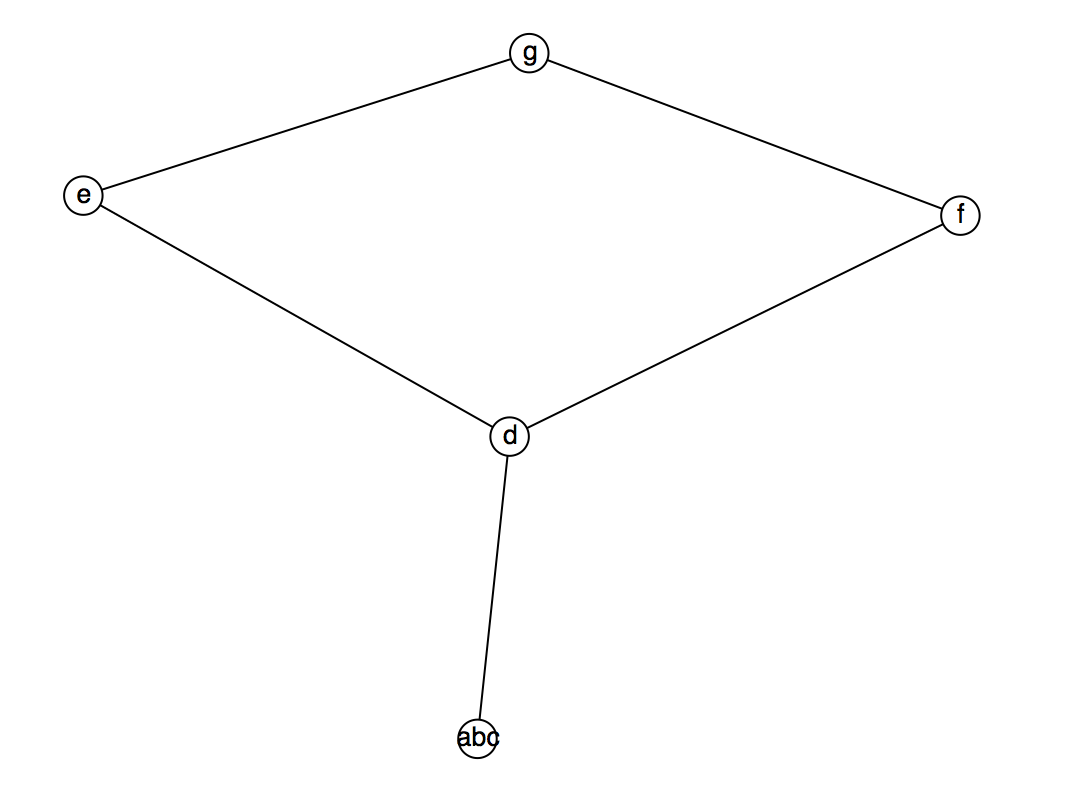}
\caption{Step 1 of reducing graph size\cite{bib6}\newline}
\label{lab:5}
\end{figure}

\begin{figure}
\includegraphics[scale=0.3]{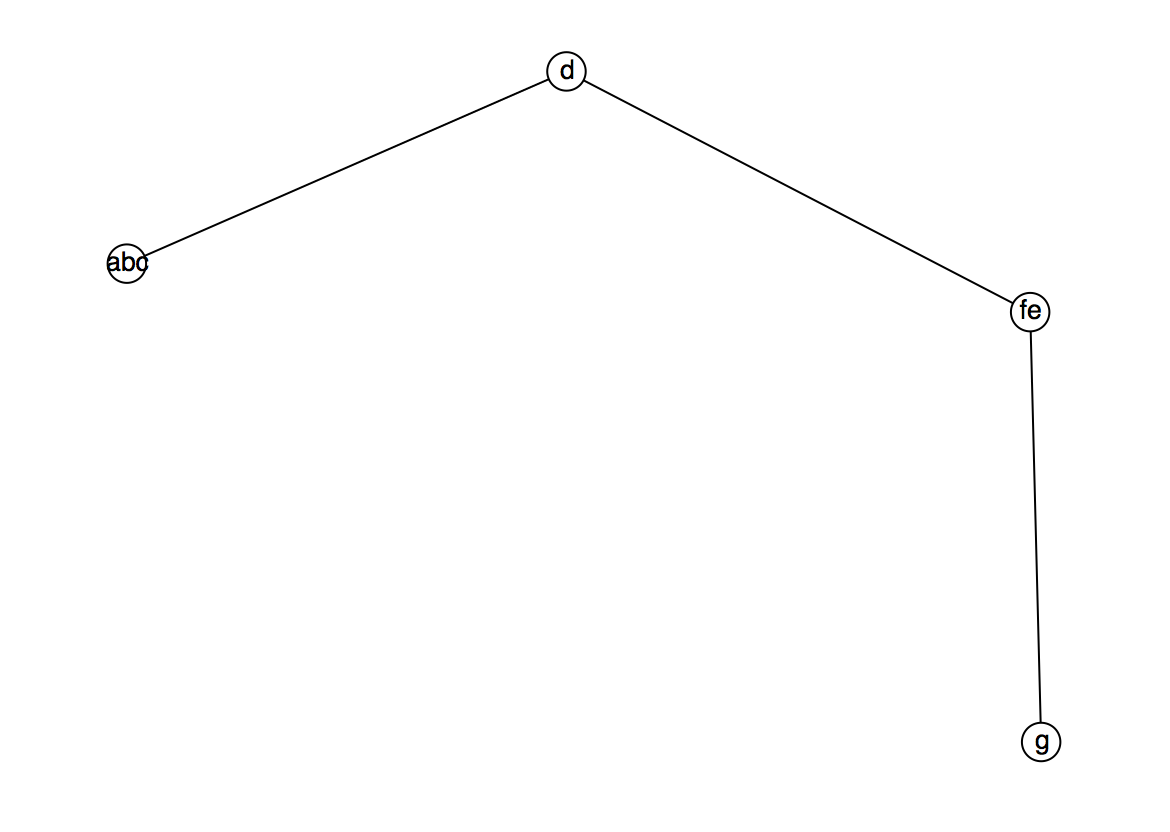}
\caption{Step 2 of reducing graph size\cite{bib6}\newline}
\label{lab:6}
\end{figure}

Module $M$ of graph $G(V,E)$ is a subset of vertices, where all of them have the same neighbors outside the set. For example, on the figure \ref{lab:1} vertices $a,b,c$ is a module, because each has one common neighbor vertex $d$. Also vertices $f$ and $e$ construct a module with common neighbors $d$ and $g$. As it seen from example, inside module vertices can be connected and/or disconnected. It results in three types of modules: parallel, series and prime. The first one describes module, where all of his vertices are disconnected, so it is basically an independent set. Whereas the second one is characterized by connected vertices. Finally, the third one relates to the set in which not all vertices are connected.\\
The modular decomposition technique suggest reducing module to one vertex with some changed quality, for example it can be weight, label or color depends on problem, and after finding all modules and reducing them, we get graph with less vertices.\\
So first, we take module $a,b,c$ and make one vertex $abc$ figure \ref{lab:5}, than reduce module $e,f$ to vertex $ef$. As result the input graph has $4$ vertices figure \ref{lab:6}.\\
Tedder et. al. proposed a linear algorithm for constructing modular decomposition tree in which the root represents input graph, its children represent strong modules (modules which do not overlap each other), after that each model decomposes to its strong module and so on, leaves of this tree are vertices of the input graph. An example of such tree shown on Diagram 1.\\

\begin{diagram}
\Tree [.Prime [.Series [.a ] [.b ] [.c ]] [.d ] [.Parallel [.e ] [.f ]] [.g ]]
\caption{Example of MD tree for graph from example 1}
\label{lab:2}
\end{diagram}
\section{Maximum clique problem}
Clique $C$ of a graph $G(V,C)$ is a subset of vertices which all of them are connected to each other.\\
Maximum clique (MC) is a clique which has a maximum size or weight, if there are weights to vertices.\\
MC problem is the NP-hard problem so why increasing the size of the input graph leads us to increasing computational time of any exact solver exponentially.
\section{The maximum clique solver based on the modular decomposition tree}
Our algorithm takes the modular decomposition tree as input data and recursively as depth-first search compute maximum clique on each level, as it solves all its children. See the pseudo code:
\begin{algorithmic}[H]
\Function{$solve$}{$node$}
	\If {$node$ $is$ $leave$}
		\State\Return $this$ $node$ $with$ $it's$ $weight$
	\EndIf
	
	\If {$node$ $has$ $type$ $parallel$}
		\State\Return $max(solve(children))$
	\EndIf
	
	\If {$node$ $has$ $type$ $series$}
		\State\Return $sum(solve(children))$
	\EndIf
	
	\If {$node$ $has$ $type$ $prime$}
		\State $subgraph = create-subgraph(children)$
		\State\Return $Ostergard(subgraph)$
	\EndIf
\EndFunction
\label{alg:1}
\end{algorithmic}
There are tree types of nodes: parallel, series and prime. When it finds node with the parallel type, it returns max of solution for its children, because they are not connected and cannot become a clique. When it finds nodes with the series type, it returns sum of children, as they are all connected. If the node $i$ type is prime   algorithm constructs a new graph, which has $n_i$ vertices ($n_i$ - the number of modules for node $i$), connects them as in input graph and gives them weights as results of calculation of MC problem and then solve it using some general algorithm for weighted maximum clique problem. In our case we use Ostergard's algorithm, because it easily implements in our algorithm, due to input data.\\
Our approach was to create such algorithm and compare it with some known solver, like Ostergard. In this article we show results against Ostergard, because it was used inside suggested approach.\\
Let's consider the graph in figure 1 and its modular decomposition tree from diagram 1. Our algorithm goes to leaf $a$ and rerurns $1$, because there is no weight to vertices, also from leafs $b$ and $c$ it returns $1$, then the algorithm goes to the node with the type series and returns $3$ as sum of $1$, $1$, $1$. At leaf $d$ and $g$ it returns $1$. At the parallel node it returns also $1$, because node $e$ and node $f$ have the same weight, if they have different weight, it will return maximum. After that at the node with type prime it constructs graph with $4$ vertices with weight $3$, $1$, $1$ and $1$ and the structure as at the figure 3, use solver and gets a solution maximum clique $a,b,c,d$ with weight is equal $4$.\\
\begin{figure}[H]
\includegraphics[width=\textwidth]{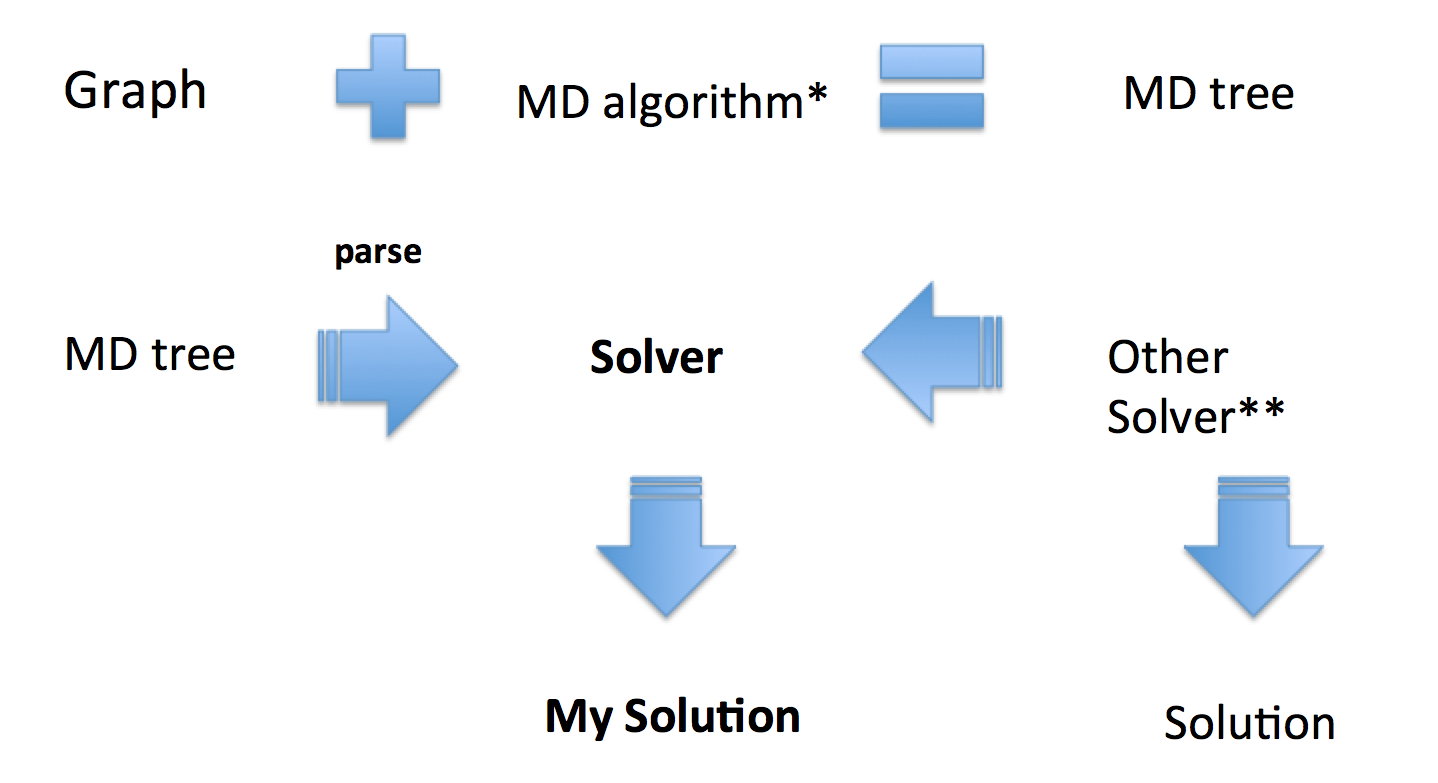}
\caption{Structure of algorithm\newline
*Marc Tedder. Derek Corneil. Michel Habib. and Christophe Paul "Simpler Linear-Time Modular Decomposition via Recursive Factorizing Permutations"\newline
** Patric R. J. Ostergard}
\label{lab:3}
\end{figure}
\section{Results on DIMACS benchmarks}
After using the algorithm of Tedder et al on DIMACS benchmarks, it was found that only few of graphs has modules, so we compare results only with them. See table 1.\\
\begin{table}
\begin{tabular}[H]{l|l|l|l|l}
Size & My algorithm (s) & MD (s) & My + MD (s) & Ostergard (s)\\
\hline
c-fat200-1&0.000118&0.068662612&0.068780612&0.001059\\
c-fat200-2&0&0.104443933&0.104443933&0.001371\\
\textbf{c-fat200-5}&\textbf{0}&\textbf{0.226667621}&\textbf{0.226667621}&\textbf{2.61894}\\
c-fat500-1&0.015956&0.147924545&0.163880545&0.002184\\
c-fat500-10&0.006224&0.680824483&0.687048483&0.011369\\
c-fat500-2&0.011647&0.216331646&0.227978646&0.003734\\
c-fat500-5&0.009128&0.324011131&0.333139131&0.006868\\
\end{tabular}
\caption{Results on DIMACS benchmarks}
\end{table}\\
As can be seen from the result table our algorithm is faster only on $c-fat200-5$, and after analysis of MD tree of this graphs, we found, that for this particular graph MD tree contains only parallel and series types of nodes. So we thought that we can create such structures to test on them.\\
\section{Algorithms for generation graphs with modules}
In this article there are two proposed algorithms to generate graphs, which in their MD tree contains only parallel and series types of nodes. It helps us to find a solution on each node easily.
\subsection{Graphs of mutual simplicity}
\begin{figure}[H]
\includegraphics[width=\textwidth]{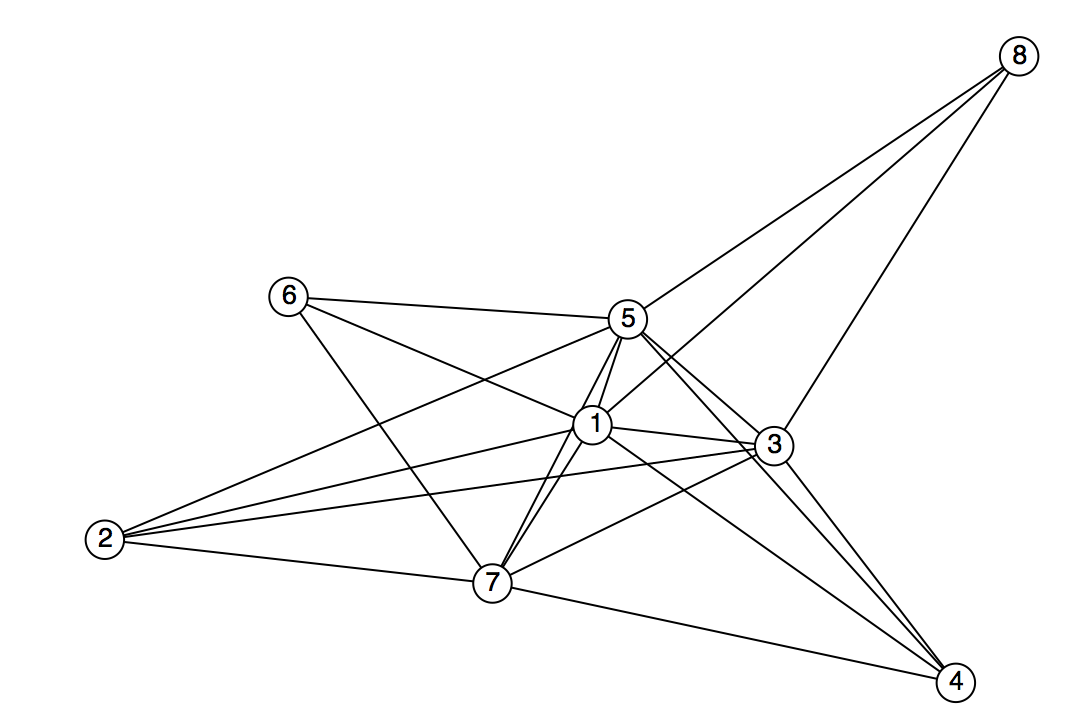}
\caption{Graph of mutual simplicity with 8 nodes}
\label{lab:4}
\end{figure}

Graph of mutual simplicity generate as follows, we connect two vertices $i$ and $j$ only if their greatest common divisor equals to $1$, for example see \ref{lab:4}.\\
For this graph you can see MD tree on Diagram 2.
\begin{diagram}
\Tree [.Series [.7 ] [.1 ] [.Parallel [.6 ] [.Series [.3 ] [.Parallel [.8 ] [.4 ] [.Series [.2 ] [.5 ] ] ] ] ] ]
\caption{MD tree for graph of mutual simplicity with 8 vertices}
\end{diagram}
\subsection{Co-Graphs}
In this case we try to build MD tree by recursively partition giving nodes to modules and randomly assign its type as parallel or series. The algorithm works as follows:
\begin{algorithmic}[H]
\Function{$partition$}{$n$}
	\State$parts = []$
	\While {$n > 0$}
		\State$p = randind(1,n)$
		\State$parts.append(p)$
		\State$n = n - p$
	\EndWhile
	\State \Return $parts$
\EndFunction

\Function{$CreateCoGraph$}{$n$}
\If {$n>1$}
	\State $parts = reverse(partition(n))$
	\If {$randint(0,1) == 0$}
		\State $graph$
		\For {$part in parts$}
			\State $subgraph = createCo-Graphs(part)$
			\State $graph.add(subgraph)$
		\EndFor
	\Else
		\State $graph$
		\For {$part in parts$}
			\State $subgraph = createCo-Graphs(part)$
			\State $graph.add(subgraph)$
		\EndFor
		\State $connect$ $all$ $subgraph$
	\EndIf
	\State \Return $graph$
\EndIf
\EndFunction
\end{algorithmic}
\section{Results on generated graphs}
\subsection{Graphs of mutual simplicity}
\begin{table}[H]
\begin{tabular}[H]{l|l|l|l|l}
Size&My algorithm(s)&MD(s)&My+MD(s)&Ostergard(s)\\
\hline
100&0&0.06906381&0.06906&0.001743\\
150&0&0.160086907&0.16009&0.004758\\
200&0&0.311571475&0.31157&0.006921\\
1000&0.00015&0.830395642&0.83055&300.001\\
1050&0.000132&2.77094437&2.77108&0.709009\\
1100&0.000149&2.962479483&2.96263&300\\
1150&0.00014&1.005088013&1.00523&300\\
1200&0.000178&1.102902136&1.10308&300\\
1250&0.000163&3.624508829&3.62467&300\\
1300&0.000237&6.191885617&6.19212&300\\
1350&0.000169&3.158776487&3.15895&300\\
1400&0.000192&0.448514811&0.44871&300\\
1450&0.000263&5.838179619&5.83844&300\\
1500&0.000273&8.211194675&8.21147&300\\
1550&0.000358&1.049262402&1.04962&300\\
1600&0.000262&1.861131225&1.86139&300\\
1650&0.000268&2.108732494&2.10900&300\\
1700&0.000283&13.137433843&13.13772&300\\
1750&0.000284&6.082156016&6.08244&300\\
1800&0.000295&2.649852882&2.65015&300\\
1850&0.00024&1.641608819&1.64185&300\\
1900&0.000331&22.834692949&22.83502&300\\
1950&0.000284&15.864648365&15.86493&300\\
2050&0.000315&19.971808447&19.97212&300\\
2100&0.000323&6.728151243&6.72847&300\\
2200&0.000331&3.152012792&3.15234&300\\
2250&0.000347&3.20890262&3.20925&300\\
2300&0.000519&26.991591521&26.99211&300\\
2350&0.000395&26.367774638&26.36817&300\\
2400&0.000368&17.514787453&17.51516&300\\
2450&0.00046&41.651295783&41.65176&300\\
\end{tabular}
\caption{Results on graphs mutual simplicity}
\end{table}
\subsection{Co-Graphs}
\begin{table}[H]
\begin{tabular}[H]{l|l|l|l|l}
Size&My algorithm(s)&MD(s)&My+MD(s)&Ostergard(s)\\
\hline
500&107 vertices 0.000119&0.599508522&0.599627522&107 vertices 5.19397\\
550&219 vertices 0.000128&1.126792247&1.126920247&219 vertices 0.420204\\
600&206 vertices 0.000123&1.613567235&1.613690235&206 vertices 0.385463\\
650&143 vertices 0.000124&1.079984486&1.080108486&37 vertices 300\\
700&197 vertices 0.000151&1.49792208&1.49807308&197 vertices 0.10201\\
750&279 vertices 0.000184&2.111751452&2.111935452&279 vertices 0.096327\\
800&92 vertices 0.00017&0.592107428&0.592277428&37 vertices 300\\
850&155 vertices 0.000172&1.634152987&1.634324987&29 vertices 300\\
900&134 vertices 0.00016&1.25222059&1.25238059&36 vertices 300\\
1000&241 vertices 0.00018&3.105684402&3.105864402&241 vertices 4.2414\\
1050&338 vertices 0.000162&4.151108273&4.151270273&116 vertices 300\\
1100&329 vertices 0.000182&4.811745176&4.811927176&329 vertices 19.4748\\
1150&408 vertices 0.000139&5.756633889&5.756772889&408 vertices 0.302615\\
1200&362 vertices 0.000165&5.614880477&5.615045477&362 vertices 2.85711\\
1250&151 vertices 0.000193&2.125359674&2.125552674&54 vertices 300\\
1300&292 vertices 0.00019&5.384282824&5.384472824&292 vertices 12.8855\\
1350&178 vertices 0.000214&2.035435868&2.035649868&63 vertices 300\\
1400&216 vertices 0.000186&2.634297357&2.634483357&32 vertices 300\\
1450&279 vertices 0.00026&5.657872455&5.658132455&35 vertices 300\\
1500&377 vertices 0.000217&9.239033114&9.239250114&377 vertices 13.7176\\
1550&238 vertices 0.000268&6.540190934&6.540458934&238 vertices 19.8803\\
1600&409 vertices 0.000262&10.753900639&10.75416264&409 vertices 48.6752\\
1650&278 vertices 0.00028&10.456971305&10.45725131&41 vertices 300\\
1700&229 vertices 0.000327&10.474212124&10.47453912&30 vertices 300\\
1800&543 vertices 0.000345&21.152246751&21.15259175&47 vertices 300\\
1850&494 vertices 0.000296&18.978203233&18.97849923&43 vertices 300\\
1900&263 vertices 0.000303&5.338127994&5.338430994&103 vertices 300\\
1950&355 vertices 0.000308&17.914419038&17.91472704&44 vertices 300\\
\end{tabular}
\caption{Results on Co-Graphs}
\end{table}
\section{Conclusion}
As can be seen from the result our algorithm works faster on graphs without prime nodes in the MD tree. It happens due to necessity to construct a subgraph and call different solver for it to calculate the maximum clique at this step. Also you can notice that the construct of MD tree takes the significant amount of calculation time, though algorithm is linear.

\section*{Funding}
This article is partially supportеd by LATNA laboratory, National Research University Higher School of Economics.


\begin{thebibliography}{}
%
%
\bibitem{bib6}
Gagneur, J., Krause, R., Bouwmeester,T. \& Casari, G. Modular decompositionof protein-protein interaction networks.Genome Biol 5, R57 (2004)
\bibitem{bib3}
Habib M., Paul C., A survey on algorithmic aspects of modular decomposition, Computer Science Review, 4: 41--59 (2010)
\bibitem{bib1}
Kuhl F. S., Crippen G. M., Friesen, D. K., A combinatorial algorithm for calculating ligand binding, Journal of Computational Chemistry,5 (1): 24-–34 (1983)
\bibitem{bib5}
Patric R.J.Östergård, A fast algorithm for the maximum clique problem, Discrete Applied Mathematics, Volume 120, Issues 1–-3, Pages 197--207(2002)
\bibitem{bib4}
Rhodes Nicholas, Willett Peter, Calvet Alain, Dunbar James B., Humblet Christine, CLIP: similarity searching of 3D databases using clique detection, Journal of Chemical Information and Computer Sciences, 43 (2): 443–-448 (2003)
\bibitem{bib2}
Tedder M., Corneil D. , Habib M., Paul C. ,Simpler linear-time modular decomposition via recursive factorizing permutations, 
35th International Colloquium on Automata, Languages and Programming, ICALP2008, Part 1, LNCS, vol. 5125, Springer, 634--64 (2008)
\end{thebibliography}


\end{document}